\documentclass[preprint,12pt]{elsarticle}

\usepackage{tikz}
\usepackage{xcolor}
\usetikzlibrary{calc, positioning}
\usepackage{graphicx} % Required for inserting imag
\usepackage{amsmath}
\usepackage{subcaption}
\usepackage{amssymb}
\usepackage{mathrsfs}
\usepackage{mathtools,leftindex,tensor,mhchem}
\usepackage{float}
\usepackage[pagebackref=false, colorlinks=true]{hyperref}
\hypersetup{
linkcolor=blue,     % color of internal links
citecolor=blue,     % color of links to bibliography
urlcolor=blue}

\journal{Physics of the Dark Universe}

\begin{document}

\begin{frontmatter}

\title{Inflation and Primordial Perturbations in Fractal Cosmology}

%% Authors
\author[inst1]{Aarav Shah}
\ead{shahaarav103@zohomail.in}

\author[inst2]{Paulo Moniz}
\ead{pmoniz@ubi.pt}

\author[inst3]{Maxim Khlopov}
\ead{khlopov@apc.in2p3.fr}

\author[inst4]{Oem Trivedi}
\ead{oem.trivedi@vanderbilt.edu}

\author[inst5,inst6]{Maxim Krasnov}
\ead{morrowindman1@mail.ru}

%% Affiliations
\affiliation[inst1]{organization={Virtual Institute of Astroparticle Physics},
  addressline={75018 Paris},
  country={France}}

\affiliation[inst2]{organization={Departamento de Física, CMA-UBI},
  addressline={Universidade da Beira Interior},
  country={Portugal}}

\affiliation[inst3]{organization={Virtual Institute of Astroparticle Physics},
  addressline={75018 Paris},
  country={France}}

\affiliation[inst4]{organization={Department of Physics and Astronomy, Vanderbilt University},
  addressline={Nashville, TN 37235},
  country={USA}}

\affiliation[inst5]{organization={National Research Nuclear University MEPhI},
  addressline={MEPhI, 115409 Moscow},
  country={Russia}}
\affiliation[inst6]{organisation={Research Institute of Physics, Southern Federal University},
addressline={344090 Rostov-on-Don},
country={Russia}}

\date{\today}

\begin{abstract}
We study inflationary dynamics within the framework of fractal cosmology, where space is characterized by an effective non-integer dimension $D$. In our work, fractal effects are sourced through thermodynamic modifications at the cosmological horizon. Using the modified Friedmann and continuity equations, we then derive the modified slow roll parameter and their evolution for linear, cubic, Starobinsky ($R+R^2$) and Natural inflationary potentials, showing that the slow roll parameters get suppressed for $D<3$. We further derive a fractal extension of the Mukhanov-Sasaki equation by introducing an effective momentum $k_{\text{eff}}$, which captures the modification of spatial Laplacian due to fractality. This leads to explicit corrections to the scalar power spectrum and the spectral index $n_s$, depending on both $D$ and a fractional length scale $L$. Confrontation with Planck 2018 data constrains the effective dimension to a best-fit range of $2.7\lesssim D \lesssim 3$ for the Starobinsky model. Furthermore, in the case of Natural Inflation, fractal corrections relax the usual requirement of super-Planckian axion decay constants, opening a phenomenologically viable parameter space inaccessible in the standard $3+1$ dimensional cosmology.

%For certain regimes, sub-horizon perturbations transition from oscillatory to exponentially suppressed behavior, revealing testable signatures of fractal inflation. {\color{blue}{you could say which are they...}} 
\end{abstract}


\end{frontmatter}

\section{Introduction}

The cosmological principle, which assumes large-scale homogeneity and isotropy of the universe, forms the cornerstone of modern cosmology \cite{Weinberg:2008cosmology}. Within this framework, the standard $\Lambda$CDM model has achieved remarkable success in explaining a wide range of observations, from the cosmic microwave background (CMB) to the distribution of galaxies on large scales \cite{Planck2018X,WMAP2003,BICEPKeck2021,DES2022,ACT2021}. To resolve the horizon, flatness, and monopole problems inherent in the standard Big Bang cosmology, the paradigm of cosmic inflation \cite{Guth:1981inflation,Linde:1982new,Albrecht:1982eternal,Lyth:2009inflation} was introduced in the 1980s. Inflation postulates a period of accelerated expansion in the very early universe, providing a natural mechanism for generating suitable primordial perturbations\cite{Planck2018X,WMAP2003,BICEPKeck2021,DES2022,ACT2021}. 
\\
\\
Despite its successes, the assumption of large-scale homogeneity has come under scrutiny. Observations of galaxy clustering, filaments, and cosmic voids \cite{Alpaslan2014,Kreckel2012,Stoica2010,Way2015,Pietronero1991,Labini2011,Joyce2000} suggest that the universe exhibits inhomogeneous, hierarchical structures across a broad range of scales. These features are naturally reminiscent of fractal geometry, raising the possibility that fractal cosmology \cite{jalalzadeh2024friedmannequationsfractalapparent,Calcagni2010,Asghari_2022,Rasouli2022,Ball2015,Modesto2009,Magliaro2009,Modesto2008,Aryal1987,Ribeiro1995,Modesto2012,Carlip2011,Modesto2010Spectral,Lauscher2005,Eichhorn2019,Baryshev2002,Trivedi_2024,bidlan2026futureripscenariosfractional,Bidlan:2025pzi,Rasouli:2024crg,Cosmai_2019,Guin_2025,doi:10.1142/S021773232140006X} could offer an alternative or complementary framework to describe cosmic structure formation \footnote{Such an interpretation would implicitly require a macroscopic value of $L$. However, in the present work we restrict our analysis to the case where $L$ is treated as a microscopic parameter.}. In this approach, the geometry of spacetime and the associated cosmological dynamics acquire scale-dependent modifications \cite{da_Silva_J_nior_2023}, potentially leaving observable imprints on inflationary dynamics and perturbations.
\\
\\
Early discussions on a possible fractal geometry of the Universe date back to the late 1980s and early 1990s. Seminal investigations include Aryal and Vilenkin’s analysis of fractal dimensions generated by inflationary scenarios \cite{Aryal1987} and the comprehensive reviews of the observational and theoretical status by Coleman \cite{coleman1992fractal}, Pietronero \cite{Pietronero1991} and Ribeiro \cite{Ribeiro1995}. Works like \cite{Rasouli2022,Rasouli:2024crg,da_Silva_J_nior_2023,Mansouri_1999} have studied quantum cosmology in the fractal setup (see for e.g. Chapter 7 of Jalalzedh \& Moniz's book \cite{Jalalzadeh:2020bqu})  Furthermore, there have been approaches where fractal inflationary models emerge from QG frameworks either through modes which have asymptotically safe gravity via running dimensions \cite{Lauscher2005,Bonanno_2002,Eichhorn2019} or through Causal Dynamical Triangulations (CDT) \cite{PhysRevLett.95.171301,Modesto2008}. Moreover, fractal inflation can also emerge from non local QG theories \cite{Calcagni2010,Calcagni2010PRL,Calcagni2012}. These approaches provide concrete realizations of an effective dimension $D$ which motivates the phenomenological framework adopted in this work (see for e.g, the work by Khorrami, Mansouri et.al \cite{khorrami1995modeluniversevariabledimension} and Mansouri \& Nasseri \cite{Mansouri_1999}.
\\
\\
While inflation has been extensively studied in the standard homogeneous setting, its formulation within a fractal framework remains relatively unexplored. The incorporation of a fractal dimension  \footnote{In the literature, the terms fractal and fractional are sometimes used interchangeably, but they have distinct conceptual meanings that must be clearly separated. In this work, “fractal” refers to the existence of an effective non-integer dimension of spacetime, denoted by $D$, which encodes scale-dependent geometric and thermodynamic deviations from standard FLRW cosmology. In contrast, the term “fractional” refers specifically to the presence of a new microscopic length scale “$L$” and to the use of fractional measures or operators that control the strength of the deformation.} into the Friedmann equations and continuity relations \cite{jalalzadeh2024friedmannequationsfractalapparent} provides a unique opportunity to test the robustness of inflationary predictions against deviations from homogeneity. Of particular interest are the implications for slow-roll dynamics, the evolution of scalar perturbations, and the scalar spectral index $n_s$, which has been precisely constrained by the 2018 Planck observations \cite{Planck2018X,WMAP2003,BICEPKeck2021,DES2022,ACT2021}. 
\\
\\
In this work, we investigate the dynamics of inflation in a fractal cosmological background \footnote{Although our framework introduces a fractional length scale $L$ and modifies the effective momentum structure (see Sec.\ref{Section 4}), this work should be clearly distinguished from fractional cosmology \cite{Garc_a_Aspeitia_2022,Gonz_lez_2023,Shchigolev_2021,Jalalzadeh_2022,leon2023cosmologyfractionalcalculusapproach,Costa_2023,Socorro_2023,Jamil_2012,socorro2023anisotropicfractionalcosmologykessence,Micolta_Riascos_2023,Calcagni_2021,doi:10.1142/S0217732326500884,article,Rasouli:2026hfy} based on fractional calculus or non-local fractional operators \cite{Oldham1974TheFC,Samko1993FractionalIA,math13223643}. In particular, we do not employ non-local fractional derivatives (as compared to say works which use Riesz's fractional derivative \cite{Jalalzadeh_2022,Bayn_2016,2010IJTP...49..270M}) in the action \cite{ElNabulsi2012,ElNabulsi2017,Jamil_2012} and model cosmology with a Orstein-Uhlenbeck like fractional differential equation \cite{El-Nabulsi:2016dsj}. Our approach is purely phenomenological and geometric, encoding derivatives from standard FLRW cosmology through an effective dimension $D$ and horizon thermodynamics. }. By deriving modified slow-roll parameters, analyzing their evolution under different potentials (Cubic, Starobinsky ($R+R^2$) \cite{Starobinsky1980} and Natural \cite{Freese_2015,Freese:1990rb,de_la_Fuente_2015}), and extending the perturbation theory to include fractal modifications, we aim to assess the viability of fractal inflationary models. In particular, we explore how the fractal dimension influences the duration of inflation, the behavior of perturbations, and the resulting predictions for the scalar spectral index \cite{Durrer2004,Piazza2013,Uggla2011,Christopherson2014,Schander2019,Martin2008,Tsujikawa2003}. Our results allow comparison with observational data, providing a means to constrain the fractal dimension and to evaluate the physical consistency of inflation within a fractal cosmology. Throughout our manuscript, we have treated the fractal dimension $D$ as a constant. This contrasts with models such as that of Abdalla, Afshordi et.al \cite{abdalla1998dynamicalapproachselfsimilaruniverse} in which, $D$ evolves dynamically from near homogeneity ($D\simeq3$) to a smaller asymptotic value, yielding a self-similar structure only in the late universe.
\\
\\
The rest of this work is structured as follows. In     Sec.\ref{Section 2}, we review the necessary fractal cosmology needed for our purposes. In Sec.\ref{Section 3}, we begin studying inflationary dynamics in a fractal setup and study the behavior of slow-roll parameters in this setup. We then further carry out our study in Sec.\ref{Section 4}, where we research perturbation theory in the inflationary phase of a fractal universe by deriving an analogous Mukhanov-Sasaki equation, studying the primordial power spectrum of scalar curvature perturbations and the scalar spectral index $n_s$ and comparing our results to the observed Planck data to obtain interesting constraints on the fractal dimension. We then conclude our work in Sec.\ref{Conclusions}.

\section{A Review Of Fractal Cosmology} \label{Section 2}
Before studying inflation in a fractal setup, it is necessary to establish the background cosmological theory in such a geometry. In fractal cosmology \cite{jalalzadeh2024friedmannequationsfractalapparent}, the large-scale structure of spacetime is described by an effective dimension $D$, introducing corrections to the standard definitions of horizon area, entropy, and expansion dynamics \cite{jalalzadeh2024friedmannequationsfractalapparent}. In what follows, we review these modifications and derive the corresponding forms of the Friedmann and continuity equations that will form the basis of our inflationary analysis.
\\
\\
In standard cosmology, spacetime is treated as a smooth, differentiable manifold of integer dimension \cite{Weinberg:2008cosmology}. However, in a fractal framework, this assumption is relaxed, and the effective dimensionality of spacetime becomes scale-dependent, reflecting the possibility that the microscopic structure of the universe may not be strictly three-dimensional. Various approaches to quantum gravity and spacetime discreteness, such as causal dynamical triangulations \cite{PhysRevLett.95.171301,Modesto2008} and asymptotic safety \cite{Lauscher2005,Bonanno_2002,Eichhorn2019}, suggest a dynamical reduction of the spectral dimension at small scales. Fractal cosmology provides a phenomenological framework for capturing such effects by modifying geometric quantities such as area, volume, and horizon entropy via an effective dimension $D\not =3$. These corrections can alter the Friedmann equations and, in turn, the inflationary dynamics.
\\
\\
To begin with, recall that the line element of the FLRW universe \cite{Weinberg:2008cosmology} is given by 
\begin{equation}
    ds^2=-dt^2+a^2(t)\left(\frac{dr^2}{1-kr^2}+r^2d\Omega^2\right),
\end{equation}
where the curvature constant $k=-1,0,+1$ addresses open, flat and closed universes respectively. In this work, we restrict ourselves to a flat universe and hence set $k=0$ throughout.
The coordinate $r$ moves with the expansion of the universe, while $\tilde{r}=a(t)r$ is the physical radial coordinate. The apparent horizon is given as
\begin{equation}\label{aparaent radius inverse hubble}
    \tilde{r}_A=a(t)r_A=\frac{1}{H},
\end{equation}
where $H=\frac{\dot{a}}{a}$ is the Hubble parameter.
\\
\\
There are several nonequivalent ways to implement a fractal modifications of cosmology. The work of G. Calcagni \cite{Calcagni2010} proposed that phenomenological fractal measures and Lebesgue-Stieltjes integration which leads to a modified scaling of area/volume. There have been works where fractal-like spectral-dimension flows emerge in discrete quantum gravity approaches such as causal dynamical triangulations and asymptotic safety \cite{Lauscher2005,Bonanno_2002,Eichhorn2019,PhysRevLett.95.171301,Modesto2008}. Furthermore, there have been thermodynamic derivations where the horizon entropy/area relations are deformed and the Friedmann equations follow from the first law on the apparent horizon (see for eg.   \cite{jalalzadeh2024friedmannequationsfractalapparent}). In this work, we adopt the thermodynamic/cosmological horizon route as our working model. 
\\
\\
The radius of the apparent cosmological horizon is then given as:
\begin{equation}
    R=(4\pi)^{\frac{D-3}{g}}\left(\frac{\tilde{r}}{L}\right)^{\frac{D}{3}}L.
\end{equation}
Consequently, the surface area $A$ and volume $V$ then generalize to:-
\begin{equation}
    A=4\pi R^2,
\end{equation}
\begin{equation}
    V=\frac{4}{3}\pi R^3,
\end{equation}
where $L$ denotes a characteristic fractional length scale associated with the underlying fractal microstructure. Throughout this work, we treat $L$ as a fixed microscopic scale of order Planck length. The factor $(4\pi)^{(D-3)/3}$ ensures that the standard FLRW relations are recovered in the limit $D\rightarrow 3$. 
\\
\\
On the apparent cosmological horizon of the FLRW cosmological model, the first law of thermodynamics is expressed as
\begin{equation} \label{Energy 1}
    dE=\mathscr{A}\Psi+Wd\mathscr{V}.
\end{equation}
where $W=\frac{-1}{2}h_{ab}T^{ab}=\frac{\rho -p}{2}$ is the work density, and $\Psi$ is given by
\begin{equation}
    \mathscr{A}\Psi =\mathscr{A}(T^{b}_{a}\partial_b \tilde{r}_A+W\partial_a\tilde{r}_A)dx^a=\mathscr{A}\psi_a dx^a.
\end{equation}
Here, $\psi_a$ denotes the projection of $T_{\mu\nu}$ on the direction normal to the boundary.
\\
\\
Now, since $dE=d(\rho\mathscr{V})=\mathscr{V}d\rho+\rho d\mathscr{V}$, eqn.\eqref{Energy 1} yields
\begin{equation}
    A\Psi=dE-Wd\mathscr{V}=\mathscr{V}d\rho.
\end{equation}
The entropy and temperature of the fractal horizon are reported (see, for e.g works like \cite{Jalalzadeh_2021,jalalzadeh2024friedmannequationsfractalapparent, Akbar_2007,Cai_2009,Debnath_2020}) to be 
\begin{equation} \label{Entropy}
    S_A=\frac{\mathscr{A}}{4G},
\end{equation}
\begin{equation} \label{temperature}
    T=\frac{1}{2\pi R}.
\end{equation}
Note that as $D\rightarrow 3$, we have $r_A\rightarrow R$ and the form of entropy given by eqn.\eqref{Entropy} reduces to that of the Bekenstein entropy form. Furthermore, the form of the temperature given by eqn.\eqref{temperature} reduces to that of the Hawking temperature form. 
\\
\\
Finally, since $T^{(fractal)}dS_A=-\mathscr{A}\Psi,$ we have,
\begin{equation}
    T^{(fractal)}dS_A=-\mathscr{V}d\rho,
\end{equation}
leading to the following equation 
\begin{equation} \label{fried}
    \frac{(4\pi)^{1-\frac{D}{3}}}{L^2}\left(\frac{L}{\tilde{r}_A}\right)^{\frac{2D}{3}}=\frac{8\pi G}{3}\rho.
\end{equation}
This expression represents a fractal modification of the standard Friedmann constraint, in which the energy density scales non-trivially with the horizon radius and the fractal dimension $D$.
\\
\\
From eqns.\eqref{fried} and \eqref{aparaent radius inverse hubble}, the generalized Friedmann equation can be expressed as
\begin{equation} \label{1}
    H^2=\left(\frac{2GL^2\rho}{3}\right)^{\frac{D}{3}-1}\frac{8\pi G}{3}\rho.
\end{equation}
Note that when $D=3$, the usual Friedmann equation of standard cosmology is recovered.
\\
\\
In this framework, energy-momentum conservation also receives a dimensional correction, leading to the fractal continuity equation,
\begin{equation}\label{2}
    \dot{\rho}+DH(\rho+p)=0.
\end{equation}
Differentiating eqn.\eqref{1}, we find
\begin{equation}\label{3}
    \dot{H}=-\frac{4D^2\pi G}{9}\left(\frac{2GL^2\rho}{3}\right)^{\frac{D}{3}-1}(\rho+p).
\end{equation}
Eqns.\eqref{1},\eqref{2} and  \eqref{3} collectively describe the cosmic evolution in a fractal spacetime of effective dimension $D$. These expressions reduce smoothly to their standard FLRW counterparts in the three-dimensional limit, confirming the internal consistency of the fractal framework.
\\
\\
This concludes our review of fractal cosmology %{\color{blue}{A nasty referee will ask for a more detailed and thorough review...}}. In the next section, we shall use equations. \eqref{1} and \eqref{2} to investigate inflation in this fractal set up.

\section{Fractal Inflation Dynamics} \label{Section 3}

%{\color{blue}{In section 3, I believe there is a footnote under equation (16), footnote number 1. It refers to the fractal universe and the apparent horizon, mentioning references [19], [53], and [54–56]. I would suggest briefly explaining what this approach is, in very simple terms, just to show the reviewer that we know what we are doing. I am being cautious here. Then, in the first paragraph of section 3, it says there is an additional dependence on the fractal dimension. I am not sure—perhaps this should be addressed later or explained in more detail here.}}

Having established the background dynamics of fractal cosmology, we now study the inflationary phase in this modified framework. In standard cosmology, inflation corresponds to a period of accelerated expansion \cite{Linde1990,Martin2014} driven by a scalar field, the inflaton, whose potential energy dominates over its kinetic term. In the fractal context, the underlying modification to the Friedmann and continuity equations introduces an additional dependence on the effective fractal dimension $D$, thereby altering the conditions for the onset and termination of inflation (by effectively changing the slow-roll parameters $\epsilon_1$ and $\epsilon_2$).
\\
\\
The energy density and pressure associated with the scalar field are given by
\begin{equation} \label{5}
    \rho_{\phi}=\frac{\dot{\phi}^2}{2}+V(\phi),
\end{equation}
and,
\begin{equation} \label{6}
    P_{\phi}=\frac{\dot{\phi}^2}{2}-V(\phi).
\end{equation}
As in the usual inflationary picture, the potential term dominates during slow roll, ensuring a quasi-de Sitter expansion.
\\
\\
The condition for accelerated expansion can be expressed as
\begin{equation}
    \frac{\ddot{a}}{a}=\dot{H}+H^2>0
\end{equation}
which implies, by eqns.\eqref{1},\eqref{3},\eqref{5},\eqref{6}, that inflation occurs when the kinetic energy of the inflaton field satisfies 
\begin{equation}\label{9}
    \dot{\phi}^2<\frac{6V(\phi)}{D^2-3}.
\end{equation}
Here $V(\phi)$ is the inflaton potential, and this inequality generalizes the standard slow-roll condition to a fractal spacetime. Eqn. \eqref{9} therefore constrains the allowed values of $D$ such that,
\begin{equation}
    D>\sqrt{3}
\end{equation}
ensuring that $\dot{\phi}$ remains real.
\\
\\
To quantify the dynamics of inflation, we define the first slow-roll parameter as
\begin{equation}
    \epsilon_1=-\frac{\dot{H}}{H^2}.
\end{equation}
A successful inflationary phase is guaranteed if and only if $|\epsilon_1|<<1$, which ensures that the expansion rate remains nearly constant. Using the modified Friedmann and continuity relations, eqns.\eqref{1},\eqref{2},\eqref{3}; we have the first slow roll parameter $\epsilon_1$ as 
\begin{equation} \label{4}
    \epsilon_1=-\frac{\dot{H}}{H^2}=\frac{D^2\dot{\phi}^2}{6\left(\frac{\dot{\phi}^2}{2}+V(\phi)\right)}.
\end{equation}
This shows that the parameter $\epsilon_1$ depends explicitly on the fractal dimension $D$; higher values of $D$ suppress the contribution of the kinetic term, thereby prolonging inflation.
\\
\\
%{\color{blue}{OK, that is interesting, I mean eq. 21 on D. Then, in equation (22) and afterwards, it talks about the presence of “D” for dimension. Be careful with what appears in those Iranian articles. Their work seems more complete and detailed. Here, “D” appears only as a constant in an equation, which seems to trivialise the concept. I would check if there are other ways to handle this, because this looks overly simplistic—but if that is indeed the case, fine.}}     
Substituting eqns.\eqref{5} and \eqref{6} into the continuity eqn.\eqref{2}, we obtain the modified Klein-Gordon equation for the inflaton field:
\begin{equation}
    \ddot{\phi}+DH\dot{\phi}+V_{,\phi}=0.
\end{equation}
The presence of $D$ in the friction term indicates that the rate of damping of the inflaton’s motion depends directly on the effective fractal dimension; a decrease in $D$  reduces the Hubble friction, allowing the inflaton field to roll down the potential more rapidly and thereby shortening the duration of inflation.
\\
\\
Differentiating the Hubble parameter once more yields,
\begin{equation}\label{Double derivative}
    \ddot{H}=-\frac{4\pi GD^2}{9}\left(\frac{2GL^2\rho}{3}\right)^{\frac{D}{3}-1 }\left[
    \begin{split}
    \left(\frac{D}{3}-1\right)\frac{(-DH\dot{\phi}^2)}{\frac{\dot{\phi}^2}{2}+V(\phi)}\\-\dot{\phi}(DH\dot{\phi}+V_{,\phi})
    \end{split}\right].
\end{equation}
Eqn.\eqref{Double derivative} encapsulates the higher-order time evolution of the expansion rate in a fractal background.
\\
\\
In the context of inflationary cosmology, it is also customary to define the second slow roll parameter as,
\begin{equation}
    \epsilon_2=\frac{\dot{\epsilon_1}}{\epsilon_1H}.
\end{equation}
A sustained slow-roll inflation requires both $|\epsilon_1|<<1$ and $|\epsilon_2|<<1$ (note that $|\epsilon_2|<<1$  is a condition for slow-roll inflation, not inflation in general). The fractal background of the universe modifies $\epsilon_2$ to,
\begin{equation} \label{7}
\begin{split}
    \epsilon_2=\frac{D\dot{\phi}^2}{3\left[\frac{\dot{\phi}^2}{2}+V(\phi)\right]}[4-D]-2D-\frac{2V_{,\phi}}{H\dot{\phi}} =2\epsilon_1[4-D]-2D-\frac{2V_{,\phi}}{H\dot{\phi}}
    \\
    =2\epsilon_1[4-D]-2D+\frac{2(DH\dot{\phi}+\ddot{\phi})}{H\dot{\phi}}= 2\epsilon_1[4-D]+\frac{2\ddot{\phi}}{H\dot{\phi}}.
\end{split}
\end{equation}
Note that in the Standard $D=3$ case, the above expression reduces to the well known form
\begin{equation}
    \epsilon_2=2\epsilon_1-6+\frac{2(3H\dot{\phi}+\ddot{\phi})}{H\dot{\phi}}=2\epsilon_1+\frac{2\ddot{\phi}}{H\dot{\phi}}.
\end{equation}
This equation indicates that both the magnitude and the evolution rate of $\epsilon_2$ are sensitive to $D$.
\\
\\
Finally, the number of e-folds passed as the inflation field evolves is given by,
\begin{equation} \label{8}
    N=\int d\phi \frac{H}{\dot{\phi}}=\int d\phi \left(\frac{2GL^2\rho}{3}\right)^{\frac{D}{6}-\frac{1}{2}}\sqrt{\frac{D}{6\epsilon_1}}\sqrt{\frac{8\pi G}{3}}.
\end{equation}
The dependence of the number of e-folds on $D$ implies that the fractal dimension effectively controls the duration of inflation.
\\
\\
\begin{figure}[H]
\centering

% ---------- epsilon1 ----------
\begin{subfigure}{\linewidth}
\centering
\includegraphics[width=\linewidth]{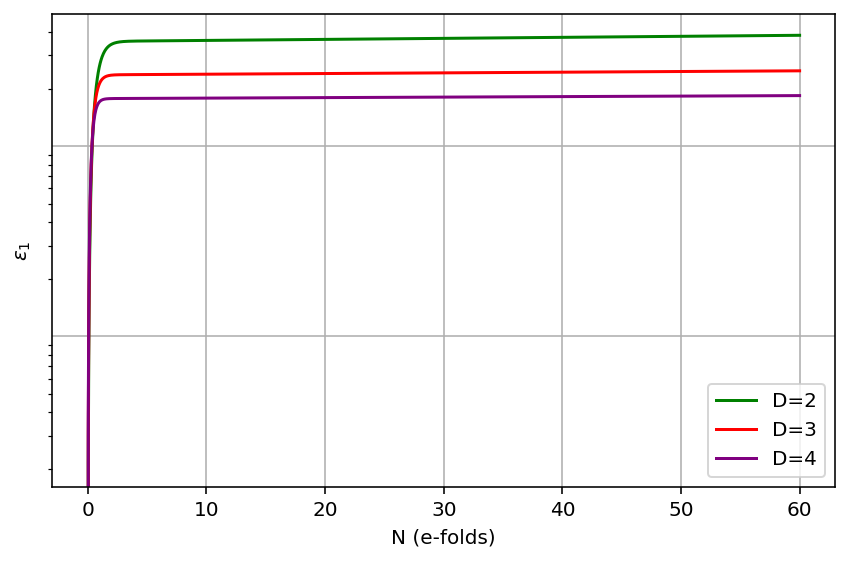}
\caption{Evolution of $\epsilon_1$ with the number of e-folds.}
\end{subfigure}
\hfill

% ---------- epsilon2 with zoom ----------
\begin{subfigure}{\linewidth}
\centering

\begin{tikzpicture}

\node (main) {
\includegraphics[width=\linewidth]{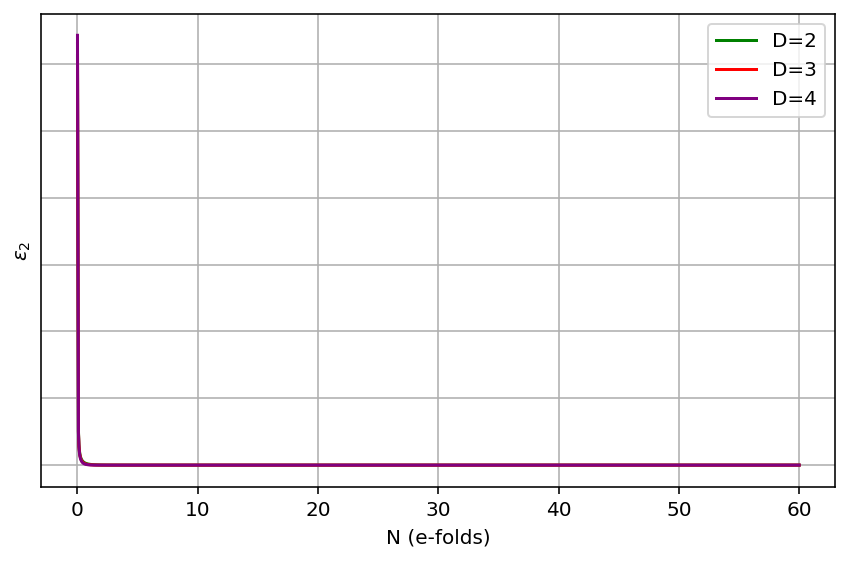}
};

% Zoom inset
\node[
draw,
thick,
fill=white,
anchor=south west
] (zoom) at ($(main.south west)+(2cm,2cm)$)
{
\includegraphics[width=0.5\linewidth]{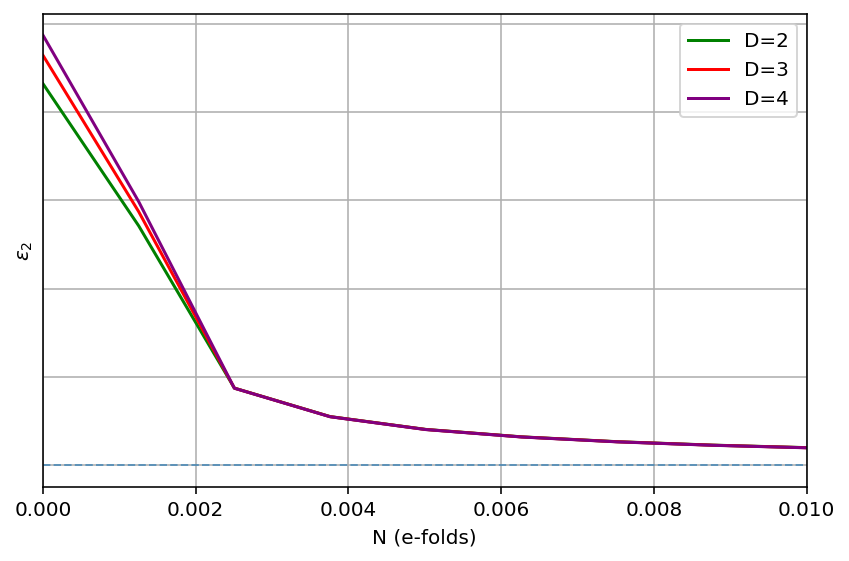}
};

% Connector lines
\draw[thick] (main.south west) ++(1.5cm,1.7cm) -- (zoom.north west);
\draw[thick] (main.south west) ++(1.5cm,1.7cm) -- (zoom.south west);

\end{tikzpicture}

\caption{Evolution of $\epsilon_2$ with a zoomed region near $N=0$.}

\end{subfigure}

\caption{
Evolution of the slow-roll parameters for the linear potential 
$V = V_0\left(1+\frac{\phi}{10}\right)$ for four values of the fractal dimension $D=2,3,4$. 
The initial conditions are $\phi=10^{-2}$ and $\dot{\phi}=0$. 
Panel (a) shows the first slow-roll parameter $\epsilon_1$, while panel (b) shows $\epsilon_2$ with an inset highlighting the early evolution for $0\le N\le 0.01$.
}

\label{Plot 1}

\end{figure}

\begin{figure}[H]
\centering

% ---------- epsilon1 ----------
\begin{subfigure}{\linewidth}
\centering
\includegraphics[width=\linewidth]{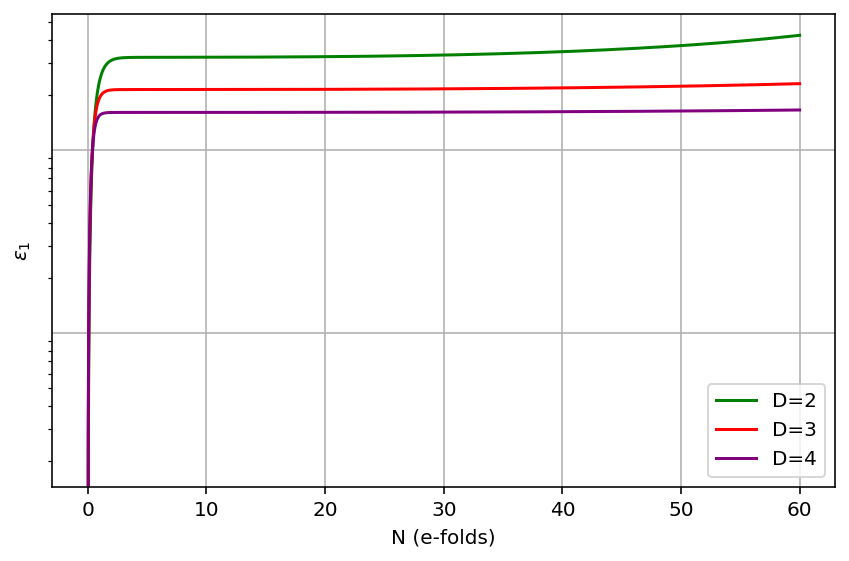}
\caption{Evolution of $\epsilon_1$ with the number of e-folds.}
\end{subfigure}
\hfill

% ---------- epsilon2 with zoom ----------
\begin{subfigure}{\linewidth}
\centering

\begin{tikzpicture}

\node (main) {
\includegraphics[width=\linewidth]{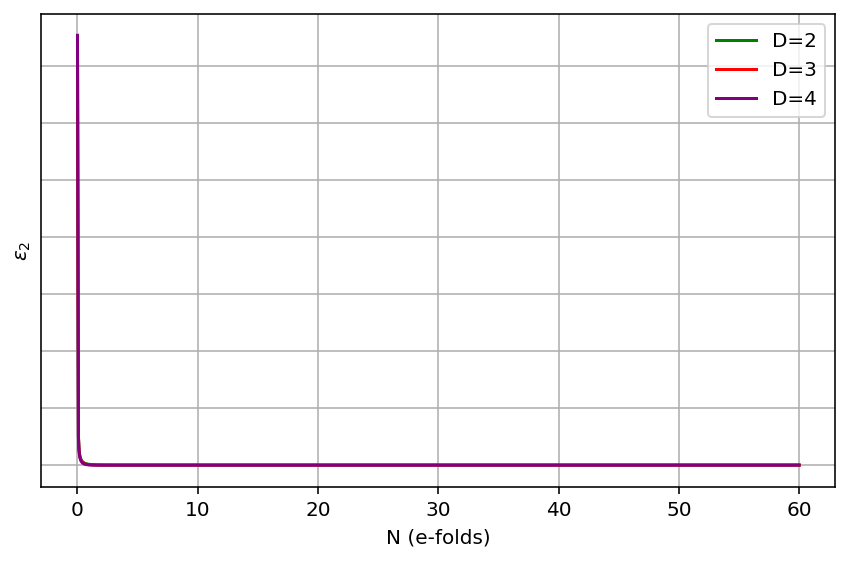}
};

% Zoom inset (positioned manually)
\node[
draw,
thick,
fill=white,
anchor=south west
] (zoom) at ($(main.south west)+(2cm,2cm)$){
  \includegraphics[width=0.5\linewidth]{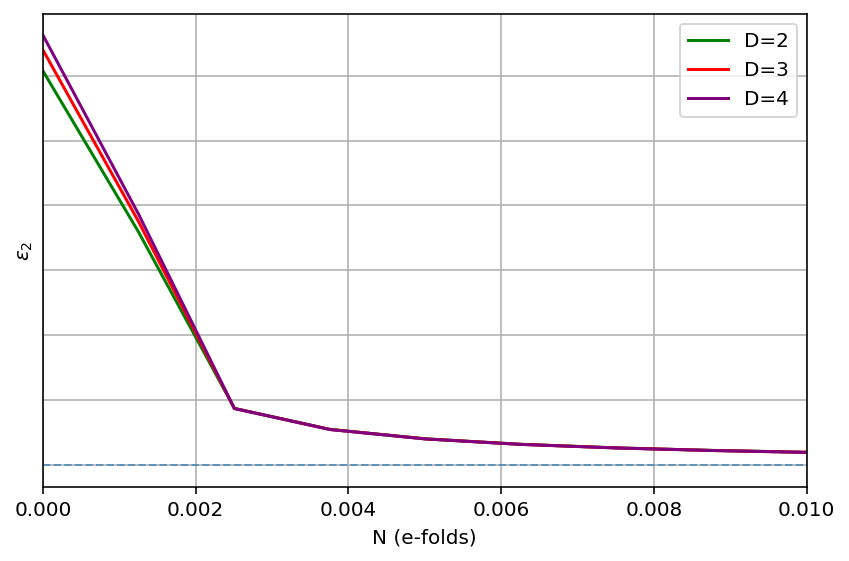}
};

% Connector lines
\draw[thick] (main.south west) ++(1.5cm,1.7cm) -- (zoom.north west);
\draw[thick] (main.south west) ++(1.5cm,1.7cm) -- (zoom.south west);

\end{tikzpicture}
\caption{Evolution of $\epsilon_2$ with a zoomed region near $N=0$.}

\end{subfigure}

    \caption{Evolution of the slow-roll parameters for the cubic potential $V = V_0\left(1+\frac{\phi^3}{10}\right)$ for four values of the fractal dimension $D=2,3,4$. The initial conditions are $\phi=10^{-2}$ and $\dot{\phi}=0$. Panel (a) shows the first slow-roll parameter $\epsilon_1$, while panel (b) shows $\epsilon_2$ with an inset highlighting the early evolution for $0\le N\le 0.01$.}
    
    \label{Plot 2}
\end{figure}
Figs.\ref{Plot 1} and \ref{Plot 2} show the evolution of the slow roll parameter $\epsilon_1$ (given by the left plot) and $\epsilon_2$ (given by the right plot) with the number of e-folds in the case of a linear and cubic potential respectively. There are two key takeaways from the left hand sided plot of figs.\ref{Plot 1} and \ref{Plot 2}. One takeaway is that as $D$ increases, it takes more number of e-folds to end inflation, in other words, $N(\epsilon_1)|_{D+1}-N(\epsilon_1)|_{D}>0$; the second is that $N(\epsilon_1)|_{D+1}-N(\epsilon_1)|_{D}$ decreases as $D$ increases. Furthermore, from the right hand sided plot of figs.\ref{Plot 1} and \ref{Plot 2}, we infer that $\epsilon_2$ very rapidly tends to $0$ as slow roll inflation starts. However, although the slow roll parameters show a dependence of the effective number of dimensions $D$, the role of $D$ in suppressing slow roll parameters subject to a monomial inflationary potential is relatively smaller than in Starobinsky and Natural inflation (discussed in later part of this manuscript). This leads us towards the direction that monomial potentials, showing less sensitivity to $D$, are more “preferred” in the fractal case than in standard $\Lambda$CDM models, matching the recent observational results of the latest CMB and BAO data \cite{balkenhol2025inflationend2025constraints}.   
\\
\\
Having analyzed monomial potentials such as cubic and linear, we now turn our attention to the Starobinsky ($R+R^2)$ model of inflation \cite{Starobinsky1980}, which arises from a purely geometric modification of gravity through the inclusion of an $R^2$ term in the action. Since this model is both theoretically well-motivated and tightly constrained by current CMB data, it provides an ideal benchmark for testing the robustness of the fractal framework against high-precision cosmological observations.
\\
\\
The Starobinsky model is formulated as a higher curvature extension of general relativity, in which the gravitational action is augmented by a quadratic correction in the Ricci scalar and can be written in the general $f(R)$ form as follows.
\begin{equation}
    f(R)=R+\frac{R^2}{6M}.
\end{equation}
where $M$ is the mass parameter. The term $\frac{R^2}{6M}$, acts like an effective “repulsive gravity,” producing accelerated expansion, without the need of a scalar field. 
\\
\\
This was originally proposed by Starobinsky \cite{Starobinsky1980}, long before the modern development of scalar-field inflation. In the early universe, the curvature $R$ becomes extremely large and hence, such corrections naturally become relevant. 
\\
\\
The action of such a theory in then reads
\begin{equation}
    S=\frac{M_{Pl}^2}{2}\int d^{D+1}x\sqrt{-g}\left(R+\frac{R^2}{6M^2}\right).
\end{equation}
In this form, the theory is defined in the Jordan frame, where the higher curvature degrees of freedom are encoded directly in the gravitational sector. It is, however, often convenient to recast the theory into the dynamically equivalent Einstein frame through a conformal transformation, whereby the effects of the quadratic curvature correction are mapped onto a scalar degree of freedom minimally coupled to gravity.
\\
\\
Upon performing the conformal transformation and redefining the scalar field to obtain a canonical kinetic term, the resulting Einstein frame scalar potential takes the form
\begin{equation}
    V(\phi)=\frac{3}{4}M^2M_{Pl}^2\left(1-e^{-\sqrt{\frac{D-1}{D}}\frac{\phi}{M_{Pl}}}\right)^2.
\end{equation}
The first slow-roll parameter, $\epsilon_1$, which governs the departure from exact de Sitter expansion and controls the end of inflation, is then given by
\begin{equation}
    \epsilon_1\simeq\frac{2(D-1)}{D}\frac{x^2}{(1-x)^2}\left(\frac{2GL^2\rho}{3}\right)^{1-\frac{D}{3}}.
\end{equation}
Here, the dimensionless variable $x$ is defined as $x=e^{-\sqrt{\frac{D-1}{D}}\phi/M_{Pl}}$, which conveniently parametrizes the evolution of the inflaton field in the Einstein frame. During slow-roll inflation one has $x<<1$, corresponding to large field values of $\phi$ in which regime the potential approaches a nearly flat plateau and ensures prolonged quasi de Sitter expansion.
\\
\\
The total number of e-folds accumulated between horizon exit and the end of inflation is given, under the slow-roll approximation, by
\begin{equation}
    N\simeq \frac{D}{2(D-1)x}\left(\frac{2GL^2\rho}{3}\right)^{\frac{D}{3}-1}.
\end{equation}
Hence, in terms of $N$, $\epsilon_1$ is now given by
\begin{equation}\label{slow roll staro 1}
    \epsilon_1=\frac{D}{2(D-1)N^2}\left(\frac{2GL^2\rho}{3}\right)^{1-\frac{D}{3}},
\end{equation}
and $\epsilon_2$ assumes the form
\begin{equation}\label{slow roll staro 2}
\begin{split}
       \epsilon_2=\frac{4x}{(1-x)^2}\frac{D-1}{D}\left(\frac{2GL^2\rho}{3}\right)^{1-\frac{D}{3}}+\left(\frac{D}{3}-1\right)\frac{DH\dot{\phi}^2}{V(\phi)}\\
       =\frac{2}{N}+\left(\frac{D}{3}-1\right)\frac{DH\dot{\phi}^2}{V(\phi)}.
\end{split}
\end{equation}
Eqn.\eqref{slow roll staro 1} shows that at a fixed number of e-folds, increasing $D$ decreases the value of the first slow roll parameter and thus delays inflation. However, at a fixed value of the exponential $x$, increasing $D$ increases the value of $\epsilon_1$. Furthermore, eqn.\eqref{slow roll staro 2} shows that for $D<3$ and $\dot{\phi}<0$, increasing the value of $D$ decreases $|\epsilon_2|$ while for $D>3$ and $\dot{\phi}<0$, increasing the value of $D$ increases $|\epsilon_2|$.
\\
\\
In the fractal framework, the slow-roll parameter $\epsilon_1$ receives an additional multiplicative suppression $\frac{D-1}{D}\left(\frac{2GL^2\rho}{3}\right)^{\frac{D}{3}-1}$ controlled by the effective dimension $D$, as shown in eqn.\eqref{slow roll staro 1}. For $\frac{2GL^2\rho}{3}<1$ (which is quite realistic for microscopic values of $L$) and $D<3$, this gravitational correction reduces $\epsilon_1$ and $|\epsilon_2|$ independently of the flattening of the inflationary potential. As a result, the unique phenomenological advantage of plateau models such as Starobinsky inflation becomes less pronounced: slow roll can be achieved without relying exclusively on extreme potential flatness. This leads to a partial degeneracy between potential-driven and geometry driven suppression mechanisms, rendering plateau models less uniquely preferred, though still fully consistent within the fractal cosmology framework. This particular result goes in hand with the recent CMB and BAO data \cite{balkenhol2025inflationend2025constraints}, providing a fractal perspective to the matter.
\\
\\
Note that, since the Higgs inflationary model \cite{chowdhury2025higgsinflationparticlefactory} is also a plateau model like Starobinsky, we similarly expect that Higgs inflation is also less preferred in a fractal universe, once again going in hand with the recent CMB and BAO data \cite{balkenhol2025inflationend2025constraints}. A full analysis regarding the RG flow of $\lambda$ \cite{SHER1989273} and the subsequent reheating \cite{shah2025particleproductionhiggsreheating} will be carried forward in future works.
\\
\\
We now turn to a qualitatively different class of inflationary scenarios, namely Natural Inflation \cite{Freese_2015,Freese:1990rb,de_la_Fuente_2015}. In contrast to the Starobinsky potential, which originates from higher curvature corrections to gravity, Natural Inflation is driven by a pseudo Nambu Goldstone boson \cite{Weinberg1972} whose flat potential is protected by an underlying shift symmetry. 
\begin{equation}
    \phi\rightarrow \phi +\text{constant}.
\end{equation}
Let $f$ be the axion decay constant, which serves to measure the extent of symmetry breaking and let $\Lambda$ set the energy scale for the symmetry breaking. The scalar potential for Natural Inflation is then given by a periodic form arising from the spontaneous breaking of a global symmetry, and is expressed as
\begin{equation}
    V(\phi)=\Lambda^4\left[1+\text{cos}\left(\frac{\phi}{f}\right)\right],
\end{equation}
which is invariant under $\phi \rightarrow \phi +2\pi f$.
\\
\\
The slow roll parameters are then found to be
\begin{equation}
    \epsilon_1=\frac{M_{Pl}^2}{2f^2}\left(\frac{2GL^2\rho}{3}\right)^{1-\frac{D}{3}}\frac{sin(\phi/f)}{(1+cos(\phi/f))^2},
\end{equation}
\begin{equation}
    \epsilon_2=-\frac{M_{Pl}^2}{f^2}\left(\frac{2GL^2\rho}{3}\right)^{1-\frac{D}{3}}\frac{cos(\phi/f)}{1+cos(\phi/f)},
\end{equation}
with the number of e-folds given as
\begin{equation}
    N\simeq \frac{f^2}{M_{Pl}^2}\left(\frac{2GL^2\rho}{3}\right)^{\frac{D}{3}-1}\text{ln}\left[\frac{sin(\phi_{end}/2f)}{sin(\phi/2f)}\right].
\end{equation}
Hence, in the natural inflationary case, the only modification sourced by the fractal dimension $D$ is that of factor $\left(\frac{2GL^2\rho}{3}\right)^{1-\frac{D}{3}}$.
\\
\\
To conclude, we have investigated slow-roll parameters of inflation in the fractal set-up, as given by eqns.\eqref{4} and \eqref{7}, and their evolution with the number of e-folds as sketched in figs.\ref{Plot 1} and \ref{Plot 2}. Note that when $D=3$, eqns.\eqref{4},\eqref{7} and \eqref{8} reduce to the expressions as given by standard cosmology.
\\
\\
We shall now derive the analogous Mukhanov-Sasaki equation in the fractal setup and investigate the role of perturbations \cite{Riotto2002,Bardeen1980,Weinberg2003}.

\section{Perturbation Theory and Connecting with Observations} \label{Section 4}

Having analyzed the background dynamics of fractal inflation, we now extend our discussion to linear perturbations of the inflaton field and the metric \cite{Riotto2002,Bardeen1980,Weinberg2003}. The behavior of these perturbations determines the statistical properties of primordial fluctuations that seed structure formation\cite{1970ApJ...162..815P,1986ApJ...304...15B}. In the standard inflationary scenario, these are governed by the Mukhanov-Sasaki equation \cite{Riotto2002,Bardeen1980,Weinberg2003}. Here, we generalize this equation to the fractal cosmology framework and identify the resulting modifications to the primordial power spectrum and spectral index.
\\
\\
In conventional inflationary perturbation theory \cite{Riotto2002,Bardeen1980,Weinberg2003}, the Fourier mode $f$ of the curvature perturbation obeys
\begin{equation}
    f''+\left(-\nabla^2_{r}-\frac{z''}{z}\right)f=0,
\end{equation}
where $r$ is the co-moving coordinate, the prime denotes differentiation with respect to conformal time $\eta$, and
 \begin{equation}
     f=a\delta\phi,
 \end{equation} 
 \begin{equation}
     z=\frac{a\dot{\phi}}{H}.
 \end{equation} 
 Performing a Fourier transform on this equation, one arrives at the following Mukhanov-Sasaki equation \cite{Mukhanov1985,Sasaki1986,Mukhanov1992} 
\begin{equation}\label{MS1}
    f''+\left(k^2-\frac{z''}{z}\right)f=0.
\end{equation}
At horizon crossing, the physical wavelength satisfies $k=aH=\frac{1}{r}=\frac{a}{\tilde{r}}$.We conjecture that in the fractal setup, the effective physical radius $R$ replaces the physical radius $\tilde{r}$ and hence, $k=\frac{a}{R}$ or,
\begin{equation}
    R=\frac{a}{k}.
\end{equation}
%In a non-fractal cosmology, the apparent horizon is given as $\tilde{r}=\frac{a}{aH}=\frac{a}{k}$. Hence, we conjecture that, when we consider a more general fractal setup, the effective apparent horizon satisfies $R=\frac{a}{k}$. 
In particular, in a fractal setup, the operator $\nabla_R$ is promoted to $\frac{ik}{a}$ in momentum representation rather than the operator $\nabla_{\tilde{r}}$ being promoted to $\frac{ik}{a}$ . 
\\
\\
The Laplacian \footnote{ There exist other works in literature who have, in a similar spirit, studied the effect of fractional modifications to the Laplacian and d'Alembertian. The work by  El-Nabulsi \& Anukool \cite{ELNABULSI2023113097} introduces functionality through Dunkl fractional Laplacian operators. Furthermore, Calcagni and Rachwa\l \space  have explored UV complete quantum field theories with a fractional d'Alembertian \cite{Calcagni_2023}} then transforms as
\begin{equation}
    \nabla_r^2=a^2\left[\frac{\partial^2R}{\partial \tilde{r}^2}\frac{\partial}{\partial R}+\left(\frac{\partial R}{\partial \tilde{r}}\right)^2\nabla^2_{R}\right]
\end{equation}
leading to,
\begin{equation}
\begin{split}
            \nabla_r^2f=
            \frac{D}{3}k^{2}a^{2-\frac{6}{D}}(4\pi)^{\frac{D-3}{3D}}L^{\frac{3}{D}-1}\\ 
            \left[
            \left(\frac{D}{3}-1\right)(4\pi)^{\frac{D-3}{3D}}L^{\frac{3}{D}-1}k^{2(\frac{3}{D}-1)}
            +k^{\frac{3}{D}-1}a^{\frac{3}{D}-1}
            \right]f.
\end{split}
\end{equation}
Substituting this result into the equation.\eqref{MS1} gives the fractal Mukhanov-Sasaki equation,
\begin{equation}
\begin{split}
    f''+ \\
   \left\{
   \begin{split}\frac{D}{3}k^{2}a^{2-\frac{6}{D}}(4\pi)^{\frac{D-3}{3D}}L^{\frac{3}{D}-1}\left[\left(\frac{D}{3}-1\right)(4\pi)^{\frac{D-3}{3D}}L^{\frac{3}{D}-1}k^{2(\frac{3}{D}-1)}+k^{\frac{3}{D}-1}a^{\frac{3}{D}-1}\right]\\
   -\frac{z''}{z}
   \end{split}\right\}f=0.
\end{split}
\end{equation}
This equation can be viewed as that of a harmonic oscillator with a modified, time-dependent frequency given by,

\begin{equation}
\begin{split}
    \omega^2=
    \frac{D}{3}k^{2}a^{2-\frac{6}{D}}(4\pi)^{\frac{D-3}{3D}}L^{\frac{3}{D}-1}\\
    \left[\left(\frac{D}{3}-1\right)(4\pi)^{\frac{D-3}{3D}}L^{\frac{3}{D}-1}k^{2(\frac{3}{D}-1)}+k^{\frac{3}{D}-1}a^{\frac{3}{D}-1}\right]
    -\frac{z''}{z}.
\end{split}
\end{equation}

We shall define an “effective momentum” as,
\begin{equation}
\begin{split}
    k_{\text{eff}}^2=\frac{D}{3}k^{2}a^{2-\frac{6}{D}}(4\pi)^{\frac{D-3}{3D}}L^{\frac{3}{D}-1}\\
    \left[\left(\frac{D}{3}-1\right)(4\pi)^{\frac{D-3}{3D}}L^{\frac{3}{D}-1}k^{2(\frac{3}{D}-1)}+k^{\frac{3}{D}-1}a^{\frac{3}{D}-1}\right].
\end{split}
\end{equation}
%Fig.\ref{Plot 4} shows the evolution of $\frac{k_{eff}}{k}$ at the horizon $k=aH$ across various values of $D$.  As expected, when $D=3,$ we recover $K=k_{eff}$ while for $D>3$,$k_{eff}>k$ and for $D<3$, $k_{eff}<k$. The numerical constancy of $\frac{k_{eff}}{k}$ reflects the fact that the Hubble parameter $H$ is approximately a constant during inflation. Furthermore, for $D=1$, $k_{eff}$ becomes imaginary ($k_{eff}^2<0$) and the behavior of $f_k$ fundamentally changes. In particular, in the sub horizon limit ($k_{eff}\eta\rightarrow \infty$), $f_k$ becomes exponential in $||k_{eff}||$ rather than oscillatory! 
%\begin{figure}
 %   \centering
 %   \includegraphics[width=0.75\linewidth]{image3.png}
 %   \caption{Evolution of $\frac{k_{eff}}{k}$ at he horizon ( $k=aH$) across values of $D$ as $D=1,2,3,4$ }
  %  \label{Plot 4}
%\end{figure}
%{\color{blue}{Section 4 might be the most interesting, but then Figure 3 is straightforward and monotonous—just straight lines. There is no discussion, nothing that makes it visually or conceptually interesting. What is the point of including only straight lines? There is nothing else. And then, in the penultimate paragraph of section 4, there is an exclamation mark after the word “oscillatory.” I would avoid this emotional tone in a scientific article; it should be as sober and objective as possible.

%From the inflationary perspective, there is a lot missing. }}
In a de-Sitter background with constant $H$, the Mukhanov-Sassaki now reduces to
\begin{equation}
    f_{\textbf{k}}''+\left(k_{\text{eff}}(k)^2-\frac{2}{\eta}\right)f_{\textbf{k}}=0.
\end{equation}
Each Fourier mode is initialized in the modified Bunch Davies vacuum,
\begin{equation}
    \lim_{k\eta\rightarrow \infty} f_{\textbf{k}}=\frac{1}{\sqrt{2k_{\text{eff}}}}e^{-ik_{\text{eff}}\eta}.
\end{equation}
which evolves to
\begin{equation}
    f_{\textbf{k}}=\frac{1}{\sqrt{2k}}\left(1-\frac{i}{k_{\text{eff}}\eta}\right)e^{-ik_{\text{eff}}\eta}.
\end{equation}
The replacement $k\rightarrow k_{\text{eff}}$ encapsulates the geometric effect of fractality on the quantum vacuum and mode evolution.
\\
\\
Subsequently, the  power spectrum for scalar perturbations in slow roll inflation at the horizon crossing limit ($k=aH$), is therefore given by,
\begin{equation}
    P^2_{\delta\phi}(k_{\text{eff}})=\left(\frac{H(t)^2}{2\pi\dot{\phi}^2}\right)^2\Big|_{k_{\text{eff}}(k=aH)}.
\end{equation}
The scalar spectral index follows from
\begin{equation}
    n_s-1=\frac{dlnP^2_{\delta\phi}(k)}{dlnk_{\text{eff}}}=\left(-2\epsilon_1-\epsilon_2\right)\left[\frac{D+3}{2D}+\frac{\left(\frac{3}{D}-1\right)\left(\frac{D}{3}-1\right)(4\pi)^{\frac{D-3}{3D}}L^{\frac{3}{D}-1}k^{\frac{3}{D}-1}}{2\left[\left(\frac{D}{3}-1\right)(4\pi)^{\frac{D-3}{3D}}L^{\frac{3}{D}-1}k^{\frac{3}{D}-1}+a^{\frac{3}{D}-1}\right]}\right],
\end{equation}
which reduces to the standard relation $n_s-1=-2\epsilon_1-\epsilon_2$ for $D=3$.
\\
\\
The Planck 2018 measurements \cite{balkenhol2025inflationend2025constraints} yielded $n_s=0.9649\pm 0.0042$, implying $n_s<1$ \cite{Planck2018X,WMAP2003,BICEPKeck2021,DES2022,ACT2021,balkenhol2025inflationend2025constraints}. For the case $2\epsilon_1+\epsilon_2>0$, this means that
\begin{equation}
\frac{D+3}{2D}+\frac{\left(\frac{3}{D}-1\right)\left(\frac{D}{3}-1\right)(4\pi)^{\frac{D-3}{3D}}L^{\frac{3}{D}-1}k^{\frac{3}{D}-1}}{2\left[\left(\frac{D}{3}-1\right)(4\pi)^{\frac{D-3}{3D}}L^{\frac{3}{D}-1}k^{\frac{3}{D}-1}+a^{\frac{3}{D}-1}\right]}>0.
\end{equation}
Let us now define $\beta$ by 
\begin{equation}
\beta=\frac{a^{\frac{3}{D}-1}}{(4\pi)^{\frac{D-3}{3D}}L^{\frac{3}{D}-1}k^{\frac{3}{D}-1}}.
\end{equation} 
The above inequality then yields,
\begin{equation}
    D>3\left(\frac{2-\beta}{\beta+2}\right).
\end{equation}
However, for the case $2\epsilon_1+\epsilon_2<0,$ the inequality is given by,
\begin{equation}
     D<3\left(\frac{2-\beta}{\beta+2}\right).
\end{equation}
In this regime the effective momentum satisfies, $k_{\text{eff}}^2<0$ implying that $f$ evolves exponentially on sub-horizon scales rather than exhibiting the usual oscillatory behavior. Such exponential evolution is incompatible with the standard inflationary picture, where sub-horizon modes must oscillate in order to define a consistent quantum vacuum. Consequently, all values of the effective dimension satisfying 
\begin{equation}
    D<3(1-\beta)
\end{equation}
lead to an unphysical sub-horizon dynamics of scalar perturbations. Requiring the dynamics to remain physical therefore imposes $D>3(1-\beta)>0$, which immediately implies $\beta <1$. For phenomenologically reasonable values of the fractional length scale $L$ and the Hubble parameter $H$ during Starobinsky ($R+R^2$) inflation, this condition restricts the effective fractal dimension to lie close to the standard three-dimensional value. In practice, consistency with oscillatory sub-horizon modes and observational considerations suggests a best-fit range of $2.7\lesssim D\lesssim 3$ for Starobinsky ($R+R^2$) inflation .
\\
\\
\begin{table*}
\centering
\caption{Planck 2018 Data Analysis}
\label{tab:mytable}
    \begin{tabular}{||c|c|c||}
        \hline
        Potential  & Value of $D$ & Results \\
        \hline
        Starobinsky ($R+R^2)$ & $D=2.7$ & $n_s\simeq 0.961616$ \\
        \hline
        Starobinsky ($R+R^2)$ & $D=2.8$ & $n_s\simeq 0.962338$ \\
        \hline
        Starobinsky ($R+R^2)$ & $D=2.9$ & $n_s\simeq 0.963009$ \\
        \hline
        Starobinsky ($R+R^2)$ & $D=3.0$ & $n_s \simeq 0.963636$ \\
        \hline
        Natural & $D=2.5$ & $f$ now satisfies $f \gtrsim 0.77 M_{Pl}$ \\
        \hline
        Natural & $D=2.6$ & $f$ now satisfies $f \gtrsim 1.12 M_{Pl}$ \\
        \hline
        Natural & $D=2.7$ & $f$ now satisfies $f\gtrsim 1.62 M_{Pl}$\\
        \hline
        Natural & $D=2.8$ & $f$ now satisfies $f\gtrsim 2.36 M_{Pl}$\\
        \hline
        Natural & $D=2.9$ & $f$ now satisfies $f\gtrsim 3.46M_{Pl}$\\
        \hline
        Natural & $D=3.0$ & $f$ now satisfies $f\gtrsim 5.00 M_{Pl}$ \\
        \hline
    \end{tabular}
\end{table*}
A close examination of Table \ref{tab:mytable} reveals that, while the Starobinsky potential remains stable under moderate deviations from $D=3$, the most substantive phenomenological effect of the fractal framework appears in the Natural Inflation scenario. In conventional four-dimensional cosmology, compatibility with CMB measurements requires a super-Planckian decay constant, $f\gtrsim 5M_{Pl}$, which has long been regarded as a major theoretical obstacle for embedding Natural Inflation in a controlled ultraviolet completion \cite{Freese_2015,Freese:1990rb,de_la_Fuente_2015}. In the fractal setting, however, the effective rescaling of the gravitational sector significantly alters this conclusion. The dependence of the slow-roll parameters on the factor $\left(\frac{2GL^2}{3\rho}\right)^{1-\frac{D}{3}}$ has a pronounced effect on Natural Inflation. For example, at $D=2.5$, the model admits values of the decay constant smaller than the Planck mass, a regime that is inaccessible in standard four-dimensional cosmology. This result may mitigate the traditional UV tension and hints towards the possibility that Natural Inflation can be reconciled with observational data within a parameter space that is inaccessible in the standard $D=3$ framework. 
\\
\\
More broadly, this illustrates how the fractal framework can modify the phenomenological viability of inflationary models rather than simply shifting numerical predictions. In particular, the modification of the slow-roll relations changes the mapping between the parameters of the potential and observable quantities such as the scalar spectral index. As a consequence, regions of parameter space that are excluded in conventional cosmology can become compatible with observational constraints. The Natural Inflation scenario provides a clear example of this effect: the fractal modification effectively enlarges the viable parameter space by relaxing the lower bound on the decay constant $f$. In this sense, the fractal framework provides a concrete phenomenological advantage, as it allows models that are otherwise theoretically disfavored due to super-Planckian field excursions to remain compatible with current cosmological observations.
\\
\\
With this, we shall conclude our manuscript; in the following section, we summarize our key findings, emphasizing how the fractal dimension influences inflationary dynamics, perturbation evolution, and potential observational signatures.

\section{Conclusion}\label{Conclusions}

In this work, we have developed a systematic analysis of inflationary dynamics \cite{Guth:1981inflation,Linde:1982new,Albrecht:1982eternal,Lyth:2009inflation} and perturbation theory \cite{Riotto2002,Bardeen1980,Weinberg2003} in the framework of fractal cosmology , where the effective dimensionality of spacetime is characterized by a non-integer parameter ($D$), emerging from a deformation of horizon thermodynamics and fractal measures \cite{jalalzadeh2024friedmannequationsfractalapparent,Calcagni2010,Asghari_2022,Rasouli2022,Ball2015,Modesto2009,Magliaro2009,Modesto2008,Aryal1987,Ribeiro1995,Modesto2012,Carlip2011,Modesto2010Spectral,Lauscher2005,Eichhorn2019,Baryshev2002,Trivedi_2024,Bidlan:2025pzi,bidlan2026futureripscenariosfractional,Rasouli:2024crg,Cosmai_2019,Guin_2025}. The resulting Friedmann and Klein-Gordon equations acquire explicit $D$ dependent corrections, which in turn modify the slow-roll dynamics and the evolution of scalar perturbations.
\\
\\
Our work spanned four classes of inflationary models: a linear potential, a cubic potential, the Starobinsky ($R+R^2$) potential \cite{Starobinsky1980} and Natural inflationary potential \cite{Freese_2015,Freese:1990rb,de_la_Fuente_2015}. In each case, the fractal dimension $D$ enters directly into the slow-roll parameters, the number of e-folds, and the form of the Mukhanov-Sasaki equation \cite{Riotto2002,Bardeen1980,Weinberg2003}. A distinctive feature of the fractal setup is the geometry induced suppression of slow-roll parameters, which acts in addition to the suppression due to potential flatness. As a result, inflation can be achieved even in models without a highly flattened potential, provided the effective dimension $D$ lies in an appropriate range.
\\
\\
A key phenomenological finding is that this geometric suppression reduces the unique observational advantage that plateau models like Starobinsky inflation \cite{Starobinsky1980} enjoy in standard four-dimensional cosmology. In the conventional scenario, the plateau potential of the Starobinsky model produces a strong flattening of the slow-roll parameters and an excellent fit to cosmic microwave background (CMB) data \cite{Bennett2013WMAP9,Hu2002CMBReview,Planck2020ResultsVI}.In the fractal context, however, an additional suppression mechanism tied to the effective dimension partially replicates this effect, diminishing the importance of potential flatness alone. Consequently, Starobinsky inflation remains consistent with the Planck 2018 observations for values $2.7 \lesssim D\lesssim 3$, but it no longer stands out as uniquely preferred once fractal effects are included.
\\
\\
Recent observational developments independently support this perspective. The latest combined CMB and BAO analysis \cite{balkenhol2025inflationend2025constraints} reports an upward shift in the scalar spectral index $n_s$, finding $n_s=0.9782\pm0.0029$ and a $95\%$ upper limit $r<0.034$. This shifts monomial potentials with $N\sim 50$ over the Starobinsky $R^2$ and Higgs inflationary models with $N\sim 51-55$ within canonical slow roll inflation.
\\
\\
Importantly, monomial models in the fractal framework are comparatively less sensitive to the geometric effects induced by $D$. Since such potentials do not rely on extreme flatness, their inflationary dynamics remain robust under moderate fractal corrections, and their compatibility with observational bounds is less contingent on the effective dimension. In this sense, fractal cosmology naturally elevates monomial potentials to a more competitive status relative to plateau models, aligning theoretical preference with the trends seen in updated observational constraints.
\\
\\
Furthermore, Natural Inflation \cite{Freese_2015,Freese:1990rb,de_la_Fuente_2015} in fractal cosmology exhibits significant phenomenological implications. The fractal corrections induce an effective rescaling of the Planck mass, which has the consequence of relaxing the conventional lower bound on the axion decay constant $f$. This may mitigate the well-known tension associated with the requirement of super-Planckian values of $f$ in standard four-dimensional Natural Inflation \cite{Freese_2015,Freese:1990rb,de_la_Fuente_2015}.  Within the fractal framework, Planck-compatible values of the spectral index allow substantially smaller decay constants, suggesting that Natural Inflation becomes easier to reconcile with ultraviolet completions.
\\
\\
At the perturbative level, we derived a fractal extension of the Mukhanov-Sasaki equation \cite{Riotto2002,Bardeen1980,Weinberg2003} by introducing an effective momentum $k_{\mathrm{eff}}$ which encodes the modification of the spatial Laplacian induced by the fractal geometry. This leads to explicit $D$ and $L$ dependent corrections to the scalar power spectrum and to the spectral index $n_s$. By comparing our predictions with the Planck 2018 constraint on the scalar spectral index, $n_s=0.9649\pm0.0042$ \cite{balkenhol2025inflationend2025constraints}, we find that the effective fractal dimension has a best-fit range $2.7\lesssim D \lesssim 3$ for Starobinsky ($R+R^2$) inflation. In the natural inflationary case, decreasing the value of $D$ effectively decreases the lower bound of the axion decay constant $f$.
\\
\\
The fractional length scale $L$, which controls the strength of the fractal deformation, appears as a new fundamental parameter of the theory. Our analysis shows that variations in $L$ can be partially degenerate with changes in $D$, but Planck-compatible solutions require $L$ to remain close to the fundamental microscopic scale, effectively suppressing large departures from standard cosmology.
\\
\\
Overall, our analysis demonstrates that fractal cosmology \cite{jalalzadeh2024friedmannequationsfractalapparent,Calcagni2010,Asghari_2022,Rasouli2022,Ball2015,Modesto2009,Magliaro2009,Modesto2008,Aryal1987,Ribeiro1995,Modesto2012,Carlip2011,Modesto2010Spectral,Lauscher2005,Eichhorn2019,Baryshev2002,Trivedi_2024,bidlan2026futureripscenariosfractional,Bidlan:2025pzi,Rasouli:2024crg,Guin_2025,Cosmai_2019} offers a novel extension of the inflationary paradigm \cite{Guth:1981inflation,Linde:1982new,Albrecht:1982eternal,Lyth:2009inflation}. 

%There are several natural extensions of this work. One immediate direction is to study models like warm inflation \cite{Berera_1995} in the fractal setup, phases like ultra slow roll \cite{Dimopoulos_2017} and modified background cosmologies like RSII braneworld \cite{Maartens_2010,shah2025ultraslowrollphasewarm}.
%{\color{blue}{see recent papers on Barrow Cosmology or entropy and how the framework changes Friedmann's eq from thermodynamics}}

\section{Acknowledgments}

The work of M. K. was performed in Southern Federal University with financial support of grant of Russian Science Foundation № 25-07-IF. P.M. acknowledges the FCT grant UID/212/2025 CMA-UBI plus the COST Actions CA23130 (Bridging high and low energies in search of quantum gravity
(BridgeQG)) and CA23115 (Relativistic Quantum Information (RQI)). The work of O.T. was supported in part by the Vanderbilt Discovery Doctoral Fellowship.
 We are grateful to Prof. Pankaj Joshi, Meet Vyas, Aum Trivedi and peers at ICSC for their support. We also thank Prof. Luigi and Prof. Gopintah Guin for helpful comments and insightful discussions.

\bibliographystyle{elsarticle-num}
\bibliography{references}
\end{document}